\documentclass[
aps,
nofootinbib,
prd,
superscriptaddress,
tightenlines,
notitlepage,
twocolumn,
showpacs,
floatfix
]{revtex4-1}

\usepackage[T1]{fontenc}
\usepackage{amssymb}
\usepackage{graphicx}
\usepackage{amsmath}
\usepackage{xcolor}
\usepackage{tensor}
\usepackage{epstopdf}
\usepackage{multirow}
\usepackage{extarrows}
\usepackage{float}
\usepackage{graphicx,subfigure}
\usepackage{braket}
 \usepackage[normalem]{ulem}
\usepackage[colorlinks=true,
citecolor=blue,
linkcolor=red,
anchorcolor=black,
urlcolor=purple]{hyperref}
\usepackage{bm}
\usepackage{multirow}
\usepackage{esvect}
\usepackage{enumerate}

\def\al{\alpha}
\def\be{\beta}

\def\de{\delta}

\def\ka{\kappa}
\def\la{\lambda}

\def\Om{\Omega}

\def\mn{{\mu\nu}}

\def\lsim{\mathrel{\rlap{\lower4pt\hbox{\hskip1pt$\sim$}}
    \raise1pt\hbox{$<$}}}
\def\gsim{\mathrel{\rlap{\lower4pt\hbox{\hskip1pt$\sim$}}
    \raise1pt\hbox{$>$}}}
\def\sqr#1#2{{\vcenter{\vbox{\hrule height.#2pt
         \hbox{\vrule width.#2pt height#1pt \kern#1pt
         \vrule width.#2pt}
         \hrule height.#2pt}}}}

\def\prt{\partial}
\def\lrpartial{\raise 1pt\hbox{$\stackrel\leftrightarrow\partial$}}

\def\part2{\partial_\alpha \partial^\alpha}

\def\pt#1{\phantom{#1}}

\def\xx'{|\vec x -\vec x'|}

\def\b2{b^\al b_\al}

\newcommand{\beq}{\begin{equation}}
\newcommand{\eeq}{\end{equation}}
\newcommand{\bea}{\begin{eqnarray}}
\newcommand{\eea}{\end{eqnarray}}
\newcommand{\bit}{\begin{itemize}}
\newcommand{\eit}{\end{itemize}}
\newcommand{\rf}[1]{(\ref{#1})}
\newcommand\bw{\begin{widetext}}
\newcommand\ew{\end{widetext}}
\newcommand{\td}{\mathrm{d}}
\newcommand{\te}{\mathrm{e}}
\newcommand{\nn}{\nonumber}

\def\0{{\sst{(0)}}}
\def\1{{\sst{(1)}}}
\def\2{{\sst{(2)}}}
\def\3{{\sst{(3)}}}
\def\4{{\sst{(4)}}}
\def\5{{\sst{(5)}}}
\def\6{{\sst{(6)}}}
\def\7{{\sst{(7)}}}
\def\8{{\sst{(8)}}}
\def\sst#1{{\scriptscriptstyle #1}}

\begin{document}

\title{\boldmath Dynamic instability analysis for bumblebee black holes: the odd parity}

\author{Zhan-Feng Mai} \email{zhanfeng.mai@gmail.com}
\affiliation{Kavli Institute for Astronomy and Astrophysics, Peking University,
Beijing 100871, China}

\author{Rui Xu} \email{xuru@pku.edu.cn}
\affiliation{Department of Astronomy, Tsinghua University, Beijing 100084, China}
\affiliation{Kavli Institute for Astronomy and Astrophysics, Peking University,
Beijing 100871, China}

\author{Dicong Liang}

\affiliation{Kavli Institute for Astronomy and Astrophysics, Peking University,
Beijing 100871, China}

\author{Lijing Shao} \email{lshao@pku.edu.cn}

\affiliation{Kavli Institute for Astronomy and Astrophysics, Peking University,
Beijing 100871, China}
\affiliation{National Astronomical Observatories, Chinese Academy of Sciences,
Beijing 100012, China}

\begin{abstract}
Spherical black-hole (BH) solutions have been found in the bumblebee gravity where a vector field nonminimally couples to the Ricci tensor. We study dynamic (in)stability associated with the gravitational and vector perturbations of odd parity against these bumblebee BHs. {\color{black} Under the plane-wave approximation,} we find that bumblebee BHs do not suffer ghost instability, but gradient instability and tachyonic instability exist when the bumblebee charge exceeds certain values. The existence of the instabilities also depends on the nonminimal coupling constant $\xi$ that, there is a minimal value $\xi \sim 4\pi G$ with $G$ the gravitational constant for the instabilities to happen. 
The theoretical consideration for bumblebee BH stability turns out to place stronger constraints on the parameter space than those from the recent observations of supermassive BH shadows by the Event Horizon Telescope Collaboration. It is also reminiscent of Penrose's cosmic censorship conjecture since the charge of bumblebee BHs cannot be too large due to the dynamic instabilities. Specifically, for $\xi(\xi-16\pi G) > 0$, we find that the charge of a bumblebee BH cannot be larger than its mass. 
\end{abstract}

\maketitle
\flushbottom

\noindent

\allowdisplaybreaks

\section{Introduction}

A framework of Lorentz-symmetry violation using effective field theory in low-energy approximation was proposed by \citet{Kostelecky:1989jp}, which is called the Standard-Model Extension (SME). 
The SME serves to systematically study theoretical effects of all possible Lorentz-symmetry violation terms as extensions of the Standard Model of particle physics and the general relativity (GR) ~\cite{PhysRevD.69.105009, PhysRevD.80.015020, Bailey:2006fd, PhysRevD.85.096005, PhysRevD.88.096006, PhysRevD.99.056016}. One of the primary Lorentz-violating terms in the gravitational sector of the SME takes the form $s^{\mu\nu} R_{\mu\nu}$, where $R_{\mu\nu}$ denotes the Ricci tensor and $s^{\mu\nu}$ is an extra field that breaks Lorentz symmetry when acquiring a nonzero background value~\cite{PhysRevD.69.105009}. To investigate the connection of SME and specific vector-tensor gravitational theories with Lorentz-symmetry violation, an action was studied by \citet{PhysRevD.69.105009}, 
\begin{widetext}
\begin{eqnarray}\label{actionbum}
 I= \int \td^4 x \sqrt{-g} \bigg[ \frac{1}{2\kappa} R +  \frac{\xi}{2\kappa}B^\mu B^\nu R_{\mu\nu} -\frac{1}{4}B^{\mu\nu}B_{\mu\nu}  
 -V \big(B^\mu B_\mu \pm b^2\big) \bigg]\,,
\label{action1}
\end{eqnarray}
\end{widetext}
where $\kappa = 8\pi G$ is related to the Newtonian constant $G$. The theory (\ref{actionbum}) is called the bumblebee gravity. Here we use $B_{\mu}$ to denote the bumblebee vector field.
The dynamical term of the bumblebee field in the action is generated by the tensor $B_{\mu\nu} \equiv \partial_\mu B_\nu - \partial_\nu B_\mu$, analogous to Maxwell's theory of electromagnetism. 
The bumblebee theory contains a nonminimal coupling term, $\sim B^\mu B^\nu R_{\mu\nu}$, between the Ricci tensor and the bumblebee field, resembling the SME Lorentz-violating term $s^{\mu\nu} R_{\mu\nu}$. 
This term is controlled by a coordinate-independent parameter $\xi$. When $\xi = 0$ and $V$ globally vanishes, the bumblebee theory reduces to the Einstein-Maxwell theory, identifying $B_\mu$ with the four-dimensional electromagnetic potential.
Compared with the Einstein-Maxwell theory, the vector field in the bumblebee theory has a self-interaction potential, $V(B^\mu B_\mu \pm b^2)$. 
For a stable vacuum of spacetime, we require that the potential $V$ is minimized when $B^\mu = b^\mu$ and $b^\mu b_\mu = \mp b^2$, implying that the bumblebee vector field has a nonzero background for a preferred frame and it violates Lorentz symmetry in the stable vacuum, analogous to the Higgs mechanism for the Higgs scalar field. 
In general, a nonvanishing minimum of the potential $V$ is equivalent to the cosmological constant. For an unknown expression of $V$, it is an often practice to probe Lorentz-symmetry violation in the asymptotically flat spacetime under the consideration of a vanishing minimum of $V$. We thus consider
\begin{eqnarray}\label{Vcon}
    &&V(B^\mu B_\mu \pm b^2)\Big|_{B^\mu = b^\mu}  = 0 \, , \cr
    && ~ \\
    &&V'(B^\mu B_\mu \pm b^2)\Big|_{B^\mu = b^\mu}  = 0  \, , \nn
\end{eqnarray}
where $V'(x) \equiv {\td V}/{\td x}$.

For the potential with a vanishing minimum, the bumblebee theory is consistent with the SME in the weak field limit \cite{Bailey:2006fd, PhysRevD.80.015020, PhysRevD.85.096005}. 
Furthermore, to probe Lorentz-symmetry violation in the strong-field region where relativistic effects of gravity become important, the properties and associated applications to compact objects, such as black holes (BHs) and neutron stars, are thus interesting topics in the bumblebee theory. 
Recently, several spherical and slowly rotating BH solutions with bumblebee vector hair and their properties in the bumblebee theory have been studied~\cite{Bertolami:2005bh, Casana:2017jkc, Gullu:2020qzu, Izmailov:2022jon, Liu:2022dcn, Xu:2022frb, Xu:2023zjw}. 
In particular, \citet{Xu:2022frb} constructed a class of spherical BHs with nonvanishing temporal component for the bumblebee field.
These spherical bumblebee BHs only have two degrees of freedom, the mass $M$ and the vector charge $Q$. 
Specifically, there are two special analytical solutions in the bumblebee theory: 
(i) the Reissner-Nordstr\"{o}m (RN) BH when $\xi =0$, 
and (ii) a stealth Schwarzschild BH carrying a nontrivial bumblebee hair when $\xi = 2\kappa$~\cite{Xu:2022frb}. 
The current observations of the shadows of the supermassive BHs, M87$^*$ and Sgr~${\rm A}^*$ from the Event Horizon Telescope (EHT) already placed some interesting bounds on $Q/M$ for various values of $\xi$~\cite{Xu:2023xqh, Xu:2022frb}. 
These constraints show that there is still a large parameter space for the bumblebee BHs yet to be excluded observationally. 
\citet{Liang:2022gdk} illustrated that it is possible to probe the vector charge-to-mass ratio up to $Q/M\sim {\cal O}(10^{-3})$ with future observations of millihertz gravitational waves from extreme-mass-ratio inspirals, so that we might distinguish bumblebee BHs from the Schwarzschild BH.  
A recent study by \citet{Hu:2023vsg} shows that timing a radio pulsar orbiting around Sgr~${\rm A}^{*}$ in a close orbit can  as well probe a vector charge-to-mass ratio as small as $Q/M\sim {\cal O}(10^{-3})$.

Regardless of the constraints on the vector charge from observations, as one of the important intrinsic properties of BHs,  BH stability answers whether such a BH can exist in our Universe from the theoretical perspective. It might give extra constraints on the parameter space of the bumblebee BHs. In a previous study, the extended thermodynamics and associated local thermodynamic stability of bumblebee BHs have been studied by us \cite{Mai:2023ggs}. Unlike the RN BH which has one phase transition point separating locally thermodynamic stability and instability, the bumblebee BHs can have zero to two phase transition points, depending on the value of the coupling constant $\xi$ \cite{Mai:2023ggs}. 

In this work, we study the dynamic stability of bumblebee BHs.
As shown by the action~\eqref{actionbum}, the bumblebee theory can be considered as a class of vector-tensor theories. 
For the special case of a vanishing $\xi$, the bumblebee theory recovers the Einstein-Maxwell theory and the RN BH solution has been proven to be stable against generic perturbations~\cite{Moncrief:1974ng, Regge:1957td}. 
For general bumblebee BHs, we are going to focus on the gradient instability and the ghost instability. These two types of instabilities have been widely investigated for hairy BHs in various modified gravity theories, such as the $f(R)$ gravity~\cite{DeFelice:2011ka, Motohashi:2011pw}, scalar-tensor theories~\cite{Kobayashi:2012kh, Kobayashi:2014wsa, Ogawa:2015pea, Ganguly:2017ort}, vector-tensor theories~\cite{Heisenberg:2017hwb, Kase:2018voo, Kase:2018owh}, and scalar-vector-tensor theories~\cite{Gannouji:2021oqz, Kase:2023kvq}. 
It seems that compact objects suffer the gradient and ghost instabilities due to nonminimal couplings between the gravitational field and extra fields. 
Specifically, \citet{Kase:2018voo} considered vectorized BHs in the vector-tensor theory including a Horndeski-like nonminimal coupling term $G_{\mu\nu}A^\mu A^\nu$, finding that there are parameter-dependent ghost and gradient instabilities associated with the odd-parity gravitational and vector perturbations. 
In addition, BHs in the Einstein-\ae{}ther theory, which is one of the Lorentz-violating gravity theories, suffer the ghost and gradient instabilities associated with the odd-parity perturbations as well~\cite{Tsujikawa:2021typ}. 
The gradient and ghost instabilities of bumblebee BHs have not been studied yet. Our work complements to the existing knowledge of BH instabilities in vector-tensor theories and provides concrete examples with detailed numerical results on the conditions for the gradient and ghost instabilities of  bumblebee BHs.

This paper is organized as follows. In Sec.~\ref{gins}, we review the ghost and gradient instabilities briefly. In Sec.~\ref{rev}, we review the basic properties of the vectorized bumblebee BHs given by \citet{Xu:2022frb}. 
In Sec.~\ref{odd}, we first introduce the gravitational and the bumblebee vector perturbations of odd parity. From the perturbed action up to the second order, we find that bumblebee BHs have no ghost instability, and then we give the generic conditions of the gradient instability. 
In Sec.~\ref{sta}, we analyze the gradient instability in detail with the numerical BH solutions, finding that in the absence of the gradient instability,  bumblebee BHs cannot carry very large values of the bumblebee charge. 
In Sec.~\ref{con}, we summarize our investigation and give some discussions for further studies. In this paper, we adopt $(-,+,+,+)$ as the metric convention and the Planck natural units, namely $G=c=\hbar=4\pi\epsilon_0 = k_B = 1$.

\section{Gradient and ghost instabilities in a toy field theory}\label{gins}

We aim to investigate the dynamic stability of bumblebee BHs. To do this, we first generally consider a field perturbation under the BH background and study whether it will grow with time. 
Let us introduce various stabilities using a toy model~\cite{Demirboga:2021nrc, Doneva:2022ewd, Delhom:2022vae}. As a simple example, we first consider a single scalar field $\psi(t,r,\theta, \varphi)$ in four-dimensional spherical coordinate. In general, the action of a free scalar field in Minkowski spacetime is
\begin{eqnarray}
    S &=& -\frac{1}{2} \int \td^4 x \left( \partial_\mu \psi \partial^\mu \psi \right) \cr
    ~\cr
    &=& \frac{1}{2} \int r^2 \td t \td r \td \Omega  \left[\dot \psi^2 - (\bm{\nabla} \psi)^2 \right]\,,
\end{eqnarray} 
where $\dot \psi$ denotes the time derivative and $\bm{\nabla}$ denotes the gradient operator in three spatial dimensions. To investigate various instabilities, we introduce two coefficients,  $c_1$ and $c_2$, and construct the following action

\begin{equation} \label{actoy}
    S=\frac{1}{2}\int r^2 \td t \td r \td\Om  \left[c_1 \dot \psi^2 - c_2(\bm{\nabla} \psi)^2 \right] \, . 
\end{equation}
Due to the spherical symmetry, one can set
\begin{equation}
    \psi(t,r,\theta, \varphi) = \sum_{\ell, m} \frac{\phi_{\ell m}(t,r)}{r} {\rm Y}_{\ell m} (\theta, \varphi) \, ,
\end{equation}
where $\ell$ and $m$ are the total and azimuthal angular momentum numbers respectively, satisfying $\ell = 0, 1, 2, \cdots$ and $m = 0, \pm 1, \pm 2, \cdots, \pm \ell$. After integrating over the angular coordinates, the action~\eqref{actoy} yields
%
\begin{align} 
    S = \frac{1}{2}\sum_{\ell, m} \int  \td t \td r \bigg[ & c_1 |\dot \phi_{\ell m}|^2 - c_2 |\phi'_{\ell m}|^2  - \frac{c_2\ell(\ell+1)}{r^2} |\phi_{\ell m }|^2 \bigg]\, .\label{actoy1}
\end{align}
Here we have performed the integration by parts with respect to $r$.

In the following, we drop the $\ell, m$ indices of the scalar mode for simplicity. Considering a Fourier mode $\phi(t,r)=\phi_k \te^{i(\omega t-k r)}$, the dispersion relation from Eq.~\eqref{actoy1} is approximately
\begin{equation}
    \omega^2 c_1 - k^2 c_2 - \frac{c_2 \ell(\ell+1)}{r^2} = 0 .
\end{equation}
Here we have taken the high-energy limit so that the derivatives of $\phi_k, \, \omega$ and $k$ are neglected. Depending on the relative size of $k$ and $\ell/r$, there are two simple cases useful to consider, namely
\begin{align} \label{sim1}
  k \gg \ell/r: ~~\quad\quad &\omega^2 = \frac{c_2}{c_1}k^2 \, , \\
  \label{sim2}
  k \ll \ell/r: ~~\quad\quad  &\omega^2 = \frac{c_2}{c_1} \frac{\ell^2}{r^2}\, .
\end{align}
We can further define the propagating speed along the radial direction as
\begin{equation}\label{speed}
    c_r \equiv \frac{\td r}{\td t} = \frac{\omega}{k} = \sqrt{\frac{c_2}{c_1}} \,, 
\end{equation}
and the propagating speed along the angular direction as
\begin{equation}\label{speed2}
\quad c_{\Omega} \equiv \frac{r \td \theta}{\td t} = \frac{r\omega}{\ell} = \sqrt{\frac{c_2}{c_1}} \,.
\end{equation}
In this simple model (\ref{actoy}), $c_r$ and $c_\Omega$ are the same.

Obviously, $\phi$ is stable when $c_1$ and $c_2$ have the same sign, implying that all modes of $\phi$ are propagating modes. On the contrary, $\phi$ is unstable when $c_1$ and $c_2$ have different signs, giving an imaginary propagating speed, implying that $\phi$ grows exponentially with time. Furthermore, the instability is usually classified into two types. When $c_1<0$ and $c_2>0$, $\phi$ is said to have the \emph{ghost instability}. When $c_1>0$ and $c_2<0$, $\phi$ is said to have the \emph{gradient instability}. In Table~\ref{tabc}, we list case by case the conditions for ghost and gradient instabilities for different $c_1$ and $c_2$.

\begin{table}[t]
    \centering
    \caption{The conditions of ghost instability and gradient instability for the action (\ref{actoy}).}
    \renewcommand\arraystretch{1.3}
    \begin{tabular}{p{3.2cm}p{1.55cm}p{1.55cm}p{1.7cm}}
\hline\hline
Instability & \multicolumn{3}{c}{Conditions} \\
\hline
     Propagating mode & $c_1 >0$    &  $c_2 >0$  & $c_r^2\, , c_\Omega^2 >0$   \\
     Ghost instability & $c_1 <0$    &  $c_2 >0$  & $c_r^2\, , c_\Omega^2 <0$   \\
      Gradient instability & $c_1 >0$    &  $c_2 <0$  & $c_r^2\, , c_\Omega^2 <0$  \\
     Propagating mode & $c_1 <0$    &  $c_2 <0$  & $c_r^2\, , c_\Omega^2 >0$ \\
\hline
\end{tabular}
\label{tabc}
\end{table}

Before we generalize the conditions to the case of multiple scalar fields, we need to point out that the instability analysis becomes less straightforward when $c_1$ and $c_2$ depend on $r$. One can only conclude that when $c_1$ and $c_2$ have the same sign for any $r$, there are no instabilities. If $c_1$ or $c_2$ becomes negative in some intervals of $r$, the existence of ghost instability or gradient instability depends on the specific expressions of $c_1$ and $c_2$. In conclusion, for $r$-dependent $c_1$ and $c_2$, we only have a necessary condition for ghost or gradient instability: $c_1 < 0 $ or $c_2 < 0$ for some intervals of $r$, or equivalently, a sufficient condition for $\phi$ being stable: $c_1 c_2> 0$ for any $r$. 

Now we discuss the ghost and gradient instabilities generalized to a toy model with multiple scalar fields. Consider $n$ coupled scalar fields ($\psi_1, \psi_2, \cdots, \psi_n$) in spherical coordinates, which can be written in the vector form
\begin{equation}
    \vv{\psi} =  (\psi_1, \psi_2, \cdots \psi_n)^\intercal \, .
\end{equation}
Expanding with the spherical harmonics,  
\begin{equation}
    \psi_i = \sum_{\ell,  m} \frac{\phi_i (t,r)}{r} {\rm Y}_{\ell m}(\theta,\varphi)
\end{equation}
and after integrating over the angular coordinates, the action of the model reduces to
\begin{equation}
   S  = \sum_{\ell, m} \int \td  t \td r \Big({\dot {\vv{\phi}}}^\intercal \boldsymbol{K} \dot{\vv{\phi}} + {\vv{\phi}'}^\intercal \boldsymbol{G} \vv{\phi}' + \vv{\phi}^\intercal \boldsymbol{M} \vv{\phi} \Big) \, ,
\end{equation}
where $\boldsymbol{K},  \boldsymbol{G} , \boldsymbol{M} $ are $n \times n$ matrices that may depend on $r$. In addition, we assume that $\boldsymbol{M}$ depends on $\ell$. Similarly, we consider a Fourier mode $\vv \phi(t,r)=\vv{\phi}_k \te^{i(\omega t-k r)}$, and take the high-energy limit to get the dispersion relation
\begin{equation}\label{dis2}
    \det \big(\omega^2\boldsymbol{K} + k^2 \boldsymbol{G} +\boldsymbol{M} \big)=0 \, .
\end{equation}

In general, it is difficult to deduce necessary and sufficient conditions from Eq.~\eqref{dis2} for the ghost and gradient instabilities. The two simplified cases in Eqs.~\eqref{sim1} and \eqref{sim2} are useful to consider. The case where the $\ell$-terms are dropped corresponds to 
\begin{equation}
    \det \big(\omega^2\boldsymbol{K} + k^2 \boldsymbol{G}  \big)=0 \, ,
\end{equation}
from which the radial propagating speed can be solved,
\begin{equation} \label{speed1}
    c_r^2 = \lim_{k\rightarrow \infty} \frac{\omega^2}{k^2} \, .
\end{equation}
The case where $k$ is dropped corresponds to 
\begin{equation}
    \det \big(\omega^2\boldsymbol{K} +\boldsymbol{M}  \big)=0 \, ,
\end{equation}
from which the angular propagating speed can be solved,
\begin{equation} \label{speed2}
    c_\Omega^2 = \lim_{\ell\rightarrow \infty} \frac{r^2 \omega^2}{\ell^2} .
\end{equation}
A set of necessary conditions for stable propagating modes then can be obtained by requiring 
\begin{equation} \label{cond1}
    c_r^2 > 0 \, , \quad\quad c_\Omega^2 > 0 \, .
\end{equation}
Because the matrices $\boldsymbol{K},  \boldsymbol{G} , \boldsymbol{M}$ generally depend on $r$, the speeds calculated using Eqs.~\eqref{speed1} and~\eqref{speed2} are functions of $r$. Equation~\eqref {cond1} requires them to be positive everywhere in the domain of interest.   

If either $c_r^2$ or $c_\Omega^2$ is negative, then instability occurs. Analogous to the single-field model, if $c_r^2<0$ or $c_\Omega^2<0$ while $\boldsymbol{K}$ is non-positive definite, then the instability is the ghost instability. If $c_r^2<0$ while $\boldsymbol{G}$ is non-negative definite, then the instability is gradient instability. Besides the ghost instability and the gradient instability, there might be another type of instability when $c_\Omega^2<0$ and $\boldsymbol{M}$ is non-negative definite. This is called the \emph{tachyonic instability}. In the single-field model, it happens to be the same coefficient $c_2$ appearing in both $c_r$ and $c_\Omega$, so the gradient instability and the tachyonic instability occur simultaneously. In the multi-field model, the elements of the matrix $\boldsymbol{M}$ are in general not related to the matrix $\boldsymbol{G}$, so the gradient instability and the tachyonic instability can occur independently.

\section{Gradient and ghost instabilities of bumblebee BHs}

\subsection{BHs with bumblebee charge }\label{rev}

We first briefly review the bumblebee BH solutions that extend the RN BH solution, and detailed derivations can be found in Refs.~\cite{Xu:2022frb, Mai:2023ggs}. {\color{black} To obtain the covariant field equations of the bumblebee theory, we perform a variation for the action of the bumblebee theory with respect to the gravitational field $g_{\mu\nu}$ and the bumblebee vector field $B_{\mu}$~\cite{Bailey:2006fd}. We then have
\begin{equation}
    G_{\mu\nu} = \kappa T^{B}_{\mu\nu} \, , \quad \nabla_{\mu}B^{\mu \nu} = \frac{\xi}{\kappa} R^{\mu\nu}B_\mu - 2 V'B^\nu \, , 
\end{equation}
where $G_{\mu\nu} \equiv R_{\mu\nu} - \frac{1}{2}g_{\mu\nu} R$ denotes the Einstein tensor and $T^{B}_{\mu\nu}$ is the energy-momentum tensor contributed by the bumblebee field. Here, $\nabla_\alpha$ is the covariant derivative operator.} The key assumptions to obtain these bumblebee BH solutions include the following. 
\begin{itemize}
	\item The potential term $V(\cdot)$ in Eq.~\rf{action1} can be ignored because either the background bumblebee field $b_\mu$ associated with the BH solutions satisfies Eq.~\eqref{Vcon} or the potential has a characteristic length at the cosmological scale so that it plays little role at the lengthscales of BHs.
	\item The background bumblebee field has only the temporal component nonvanishing, namely $b_\mu = (b_t, 0, 0, 0)$. In principle, all spherical, static BH solutions with $b_\mu = (b_t, b_r, 0, 0)$ in the bumblebee gravity were obtained by \citet{Xu:2022frb}. We here focus on the branch of bumblebee BHs with temporal component only.
\end{itemize}

{\color{black} With these assumptions, } the field equations for the metric of bumblebee gravity and the associated {\color{black} background} bumblebee field $b_\mu$ are then \cite{Xu:2022frb}
\begin{equation}\label{eom2}
G_{\mu\nu}=\kappa T^{b}_{\mu\nu}\, , \quad\quad \nabla_\mu b^{\mu \nu }=\frac{\xi}{\kappa}R^{\mu\nu}b_\mu \, ,
\end{equation}
where $b_\mn = \prt_\mu b_\nu - \prt_\nu b_\mu$, and the energy-momentum tensor contributed by $b_\mu$ is
\begin{widetext}
\begin{eqnarray}
 T^{b}_{\mu\nu}   \equiv \left.T^{B}_{\mu\nu}\right|_{B_\mu = b_\mu} =   \frac{\xi}{2\ka} && \bigg( g_{\mu\nu} b^\al b^\be R_{\al\be} - 2 b_\mu b_\la R_\nu^{\pt\nu \la}  -  2 b_\nu b_\la R_\mu^{\pt\mu \la} - \Box_g ( b_\mu b_\nu )  - g_{\mn} \nabla_\al \nabla_\be ( b^\al b^\be ) \cr
&&~\\
&& +  \nabla_\ka \nabla_\mu \left( b^\ka b_\nu \right) + \nabla_\ka \nabla_\nu ( b_\mu b^\ka )   \bigg)   + b_{\mu \lambda}b_\nu{}^{\lambda}-\frac{1}{4}g_{\mu\nu}b^{\alpha\beta}b_{\alpha\beta} \, . \nonumber
\end{eqnarray}
\end{widetext}
Here $\Box_g \equiv \nabla_\alpha \nabla^\alpha$ is defined as the d’Alembertian in the curved spacetime.

With the spherical ansatz for the metric
\begin{eqnarray}\label{sphe1}
&&\td s^2 = -h(r)\td t^2 + \frac{\td r^2}{f(r)}+ r^2 \left(\td \theta^2 + \sin^2 \theta \td \varphi^2 \right)  ,
\end{eqnarray}
one gets three equations from Eq.~\rf{eom2} to solve for the metric functions $h(r)$, $f(r)$, and the bumblebee field component, $b_t(r)$. {\color{black} The explicit field equations under the static spherical ansatz are displayed in Appendix~\ref{app1} }. Two analytical solutions are worth mentioning.  
\begin{enumerate}[(I)]
\item When $\xi = 0$, we have
\bea
&& f=h= 1-\frac{2M}{r}+ \frac{Q^2}{r^2} , \quad b_t = \mu_{\infty} - \sqrt{\frac{2}{\kappa}}\frac{Q}{r} \, .
\label{specase1}
\eea
It is the RN solution as one expects for $\xi=0$ ($\mu_\infty$ is a constant usually taken to be zero). 
\item When $\xi = 2\kappa$, we have 
\bea
&& f=h= 1-\frac{2M}{r} \, ,~ b_t = \frac{Q}{\sqrt{2\kappa}M}\left(1-\frac{2M}{r}\right) \, .
\label{specase2}
\eea
It is an interesting case showing that the Schwarzschild metric can be accompanied by a simple but nonzero bumblebee field. 
\end{enumerate}

For other arbitrary values of $\xi$, we could not find analytical solutions. Numerical BH solutions are calculated instead. These numerical BHs have the following asymptotic behavior,
\begin{eqnarray}\label{inf}
&& h(r) \Big|_{r \to \infty} = 1- \frac{2M}{r} + \frac{\tilde h_2}{r^2} + \cdots \, , \cr
~\cr
&& f(r) \Big|_{r \to \infty} = 1- \frac{2M}{r} + \frac{\tilde f_2}{r^2} + \cdots \, , \\
~\cr
&& b_t(r) \Big|_{r \to \infty} = \mu_\infty -\sqrt{\frac{2}{\kappa}}\frac{Q}{r} + \frac{\tilde b_2}{r^2} + \cdots  \, , \nn
\end{eqnarray}
where the ADM mass $M$ and the bumblebee charge $Q$ are the only two free parameters for the solutions, and other expansion coefficients, $\mu_\infty, \, \tilde h_2, \, \tilde f_2, \, \tilde b_2, \, \cdots$, are recursively related to $M$ and $Q$.
At the event horizon $r=r_h$, the functions $h, \, f, \, b_t$ behave as
\begin{eqnarray}\label{ayho}
&& f(r) \Big|_{r \to r_h} = f_1 (r - r_h) + f_2 (r -r_h)^2  + \cdots \, , \cr
~\cr
&& h(r) \Big|_{r \to r_h} = h_1 (r - r_h) + h_2 (r -r_h)^2 + \cdots \, , \cr
~\cr
&& b_t(r) \Big|_{r \to r_h} =b_{t1}(r-r_h) + b_{t2}(r-r_h)^2 + \cdots ,
\end{eqnarray}
where the coefficients, $f_1, \, h_1, \, b_{t1}, \, f_2, \, h_2, \, b_{t2}, \, \cdots$, and the horizon radius $r_h$ can be related to $M$ and $Q$ numerically.

\begin{figure}[t]
  \begin{center}
\includegraphics[width=0.47\textwidth]{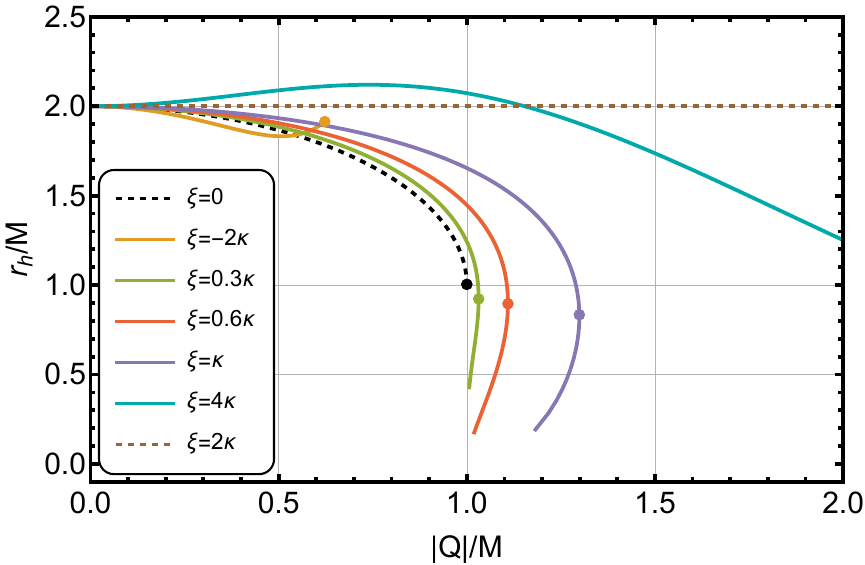}
  \caption{BH horizon $r_h$ versus BH charge $|Q|$ for different $\xi$. } 
  \label{rhQ}
  \end{center}
\end{figure}

In Fig.~\ref{rhQ}, we plot $r_h$ with respect to the bumblebee charge $Q$ for different values of $\xi$. We note that when $\xi<2\ka$, the BHs have maximal vaules for the bumblebee charge; a typical example is $Q_{\rm max} = M$ in the RN case when $\xi=0$. When $\xi=2\ka$, the numerical result shows that $Q$ becomes unrestricted, agreeing with the analytical solution in Eq.~\rf{specase2}. For $\xi>2\ka$, our numerical method can produce very large values of $Q$ with extremely small $r_h$, suggesting that $Q$ is also unrestricted when $\xi>2\ka$. But the errors of our numerical solutions increase rapidly at large $Q$ so that the solutions quickly become untrustworthy before we can extract further information. 

To figure out whether $Q$ has maximal values when $\xi>2\ka$, we have made another insufficient yet plausible attempt. We find approximate solutions around $\xi=2\ka$,
\bea
&& f = h = 1-\frac{2M}{r} - \frac{\de \xi}{2\ka} \frac{Q^2}{r^2} + O\big(\de\xi^2\big) ,
\nonumber \\
&& b_t = \frac{Q}{\sqrt{2\kappa}M}\left( 1-\frac{2M}{r} - \frac{\de \xi}{2\ka} \frac{Q^2}{r^2} + O\big(\de\xi^2\big) \right),
\eea   
where $\de\xi = \xi-2\ka$. Then the radius of horizon has an approximation 
\bea
r_h = 2M + \frac{\de\xi}{4\ka} \frac{Q^2}{M} + O\big(\de\xi^2\big).
\eea
For the approximation to work at a small enough $|\de\xi|$, we have an estimation
\bea
\frac{|\de\xi|}{4\ka} \frac{Q^2}{M} \lesssim 2M ,
\eea
which tells us that $Q/M$ is bounded at least at a small enough $|\de\xi|$ via
\bea
\frac{Q}{M} \lesssim \sqrt{\frac{8\ka}{|\de\xi|}} .
\eea
We cannot think of a reason for $Q$ being unbounded when $\de\xi$ is large if it already has a bound around $\xi=2\ka$. So we tentatively conclude that $Q$ may also have maximal values when $\xi>2\ka$. 
The corresponding boundary is currently represented by the largest values of $Q$ that we can find using our numerical code while limiting the relative error in the radius of horizon to $5\%$.

%

\subsection{Field perturbation:  odd parity}\label{odd}

With the bumblebee BH solutions in hand, we are ready to investigate the dynamical stability of bumblebee BHs. We consider a gravitational perturbation $\epsilon h_{\mu\nu}$ and a vector perturbation $\epsilon \delta b_\mu$ in the bumblebee BH background with $\epsilon \ll 1$.
Then, including the bumblebee BH background $g_{\mu\nu}$ and bumblebee vector $b_\mu$, the perturbed metric and bumblebee vector field are given by
\begin{equation}
    g_{\mu\nu} \to  g_{\mu\nu} + \epsilon h_{\mu\nu} \,  , \quad\quad    b_\mu \to  b_\mu + \epsilon \delta b_\mu \, .
\end{equation}
In general, on the spherical ansatz background, the perturbations can be decomposed into odd parity modes and even parity modes based on the following rules. 
For the gravitational perturbation $h_{\mu\nu}$, the components $h_{tt}$, $h_{tr}$ and $h_{rr}$ transform as scalars under rotation on the two-dimensional sphere of $\theta$ and $\varphi$. For example,  an arbitrary scalar function, $\Psi(t,r,\theta,\varphi)$, can be written as summation of spherical harmonic function, ${\rm Y}_{\ell m} (\theta, \varphi)$, as
\begin{equation}
    \Psi(t,r,\theta,\varphi) = \sum_{\ell, m} \Psi_{\ell m} (t,r) {\rm Y}_{\ell m}(\theta,\varphi) \, .
\end{equation}
Since the spherical harmonic function transforms as 
$${\rm Y}_{\ell m} (\pi - \theta,\varphi + \pi) = (-1)^{\ell}{\rm Y}_{\ell m}(\theta,\varphi) $$ 
under reflection on the two-dimensional sphere, one finds 
$$\Psi(t,r, \pi - \theta,\varphi + \pi) = (-1)^{\ell}\Psi (t,r,\theta,\varphi) \,.$$
It is called the even parity mode. 
On the other hand, with $a,b=\theta, \varphi$, $h_{ta}$ and $h_{ra}$ transform as vector, while $h_{ab}$  transforms as tensor under two-dimensional reflection. For the vector or tensor components, the mentioned even parity modes transform with the $(-1)^{\ell}$ factor, while the others transform with the $(-1)^{\ell+1}$ factor and are called odd parity modes~\cite{Regge:1957td}. 
Given an arbitrary vector or symmetric tensor on the two-dimensional sphere, $V_a$ and ${\cal T}_{ab}$, they can be decomposed into,
\begin{align}
     V_a =& D_a \Psi_1 + \sqrt{\gamma}\varepsilon^b{}_a D_b \Psi_2\, , \cr
     {\cal T}_{ab}=& D_a D_b \Psi_3 + \gamma_{ab} \Psi_4  + \frac{\sqrt{\gamma}} {2}\big(\varepsilon^c{}_a D_c D_b + \varepsilon^c{}_b D_c D_a\big)\Psi_5 \nn \, ,
\end{align}
where the anti-symmetric tensor $\varepsilon_{ab}$ is defined on the two-dimensional sphere with $\varepsilon_{\theta\varphi}=1$, and $\Psi_i$ are scalar functions. We denote $\gamma_{ab}$ as the induced metric on this sphere, with $\gamma$ its determinant and $D_a$ its corresponding covariant derivative. 

In this paper, we focus on the bumblebee BHs' dynamical stability associated with the odd parity perturbations. Following the above illustrations, the gravitational perturbation $h_{\mu\nu}$ of odd parity can be written as
\begin{align}
    h_{tt} &=h_{rr}=h_{rt}=0 \, ,  \nn \\
    h_{ta} &=\sum_{\ell, m}h_{0\ell m} (t,r)\sqrt{\gamma}\varepsilon_a{}^b D_b {\rm Y}_{\ell m} 
    (\theta,\varphi) \, ,  \nn \\
    h_{ra} &=\sum_{\ell, m}h_{1\ell m} (t,r)\sqrt{\gamma}\varepsilon_a{}^b D_b {\rm Y}_{\ell m} 
    (\theta,\varphi) \, , \nn \\
    h_{ab}&=\frac{1}{2} \sum_{\ell, m}h_{2\ell m} (t,r)\sqrt{\gamma} \big(\varepsilon^c{}_a D_c D_b + \varepsilon^c{}_b D_c D_a\big){\rm Y}_{\ell m}(\theta,\varphi) \nn \, .
\end{align}
The odd parity perturbation of the bumblebee vector can be written as
\begin{align}
    \delta b_t &= \delta b_r = 0 \, ,  \\
    \delta b_a &= \sum_{\ell, m} \beta_{\ell m}(t, r) \sqrt{\gamma}\varepsilon_a{}^b \partial_b {\rm Y}_{\ell m}. 
\end{align}
It is worthwhile to point out that not all perturbations are physical because of the gauge degrees of freedom. 
Note that the diffeomorphism still holds in the bumblebee theory, and one can consider an infinitesimal diffeomorphism transformation, $x^\mu \to x^\mu + \lambda^\mu$, yielding
\begin{equation}
    h'_{\mu\nu} \to h_{\mu\nu} + \nabla_\nu \lambda_\mu + \nabla_\mu \lambda_\nu \, .
\end{equation}
It indicates that one can always find four scalar functions to simplify the perturbations. Based on the decomposition rules for odd parity modes, we consider
\begin{align}
    \lambda_t &= \lambda_r =0 \, , \\
    \lambda_a &= \sum_{\ell, m}\Lambda_{\ell m} (t,r) \sqrt{\gamma}\varepsilon_a{}^b \partial_b {\rm Y}_{\ell m} \,,
\end{align}
such that the gravitational perturbations, $h_{0 \ell m}$, $h_{1 \ell m}$, and $h_{2\ell m}$, respectively transform as
\begin{eqnarray}\label{trandiff}
   &&  h_{0 \ell m} \to h_{0 \ell m} + \frac{\td \Lambda_{\ell m}}{\td t} \, ,    \cr
   ~\cr
   && h_{1\ell m} \to h_{1 \ell m} + \frac{\td \Lambda_{\ell m}}{\td r} - \frac{2}{r} \Lambda_{\ell m} \,, \\
   ~\cr
   && h_{2\ell m} \to h_{2 \ell m} + 2 \Lambda_{\ell m} \, . \nn
\end{eqnarray}
Specifically, the monopole with $\ell =0$ does not exist in odd-parity modes. As for the dipole mode with $\ell = 1$, $h_{ab}$ vanishes identically. We therefore discuss this case separately. 
For $\ell \geq 2$, we shall choose the Regge-Wheeler gauge such that $h_{2 \ell m} = 0$ and two physical components, $h_{0 \ell m}$ and $h_{1\ell m}$, are left~\cite{Regge:1957td}. 
Following the above arguments, the gravitational perturbation, together with the bumblebee vector perturbation, can be rewritten in the following matrix form,
\begin{widetext}
\begin{align}\label{hbeta}
   h_{\mu\nu} &= \sum_{\ell, m}
   \begin{pmatrix}
   0 && 0 && -  {h_0(t,r)}  \sin^{-1} \theta \, \partial_\varphi && \sin \theta \, h_0(t,r) \partial_\theta \\
   0 && 0 && - {h_1(t,r)}  \sin^{-1} \theta \, \partial_\varphi && \sin \theta \, h_1(t,r) \partial_\theta \\
   * && * && 0 && 0 \\
   * && * && 0 && 0 \\
   \end{pmatrix} {\rm Y}_{\ell m} (\theta, \varphi) \, , \\
   \delta b_\mu (t,r) &= \sum_{\ell, m} \beta(t,r)
    \begin{pmatrix}
        0\\
        0\\
        - \sin^{-1} \theta \,\partial_\varphi \\
        \sin \theta \, \partial_\theta
    \end{pmatrix} {\rm Y}_{\ell m}(\theta,\varphi) \, , \label{hbeta2}
\end{align}
%
where the ``*'' represents the symmetric part of the matrix.
For convenience, in the following, we drop the $\ell, m$ indices. With Eqs.~(\ref{hbeta}) and (\ref{hbeta2}), we expand action~\eqref{actionbum} up to ${\cal O}(\epsilon^2)$. 
Then, after performing integration with respect to $\theta$ and $\varphi$, as well as integrating by parts with respect to $t$ and $r$, the resulting perturbed action of odd parity modes reads
\begin{equation}\label{acpert}
I = I_0 +  \epsilon^2 \sum_{\ell, m} \frac{\ell(\ell+1)}{4 \kappa }I_{\rm odd} \, , 
\end{equation}
where $I_0$ denotes the on-shell action satisfying the equations of motion~\eqref{eom2}. Moreover, we have 
%
\begin{eqnarray} \label{action2}
    I_{\rm odd} =   \int \td t \td r \Bigg[  && C_1 \left(\dot h_1 - h_0' + \frac{2}{r} h_0\right)^2   + 2 ( C_2 \beta' + C_3 \beta) \left(\dot h_1 - h_0' + \frac{2}{r} h_0\right)  + C_4 \dot \beta^2 + C_5   \beta'^2    \cr
~ \cr
    && + \left(C_6 h_1^2 + C_7 h_0^2 + C_8 \beta h_0\right) + C_9 \beta^2 \Bigg]  \, ,
\end{eqnarray}
\end{widetext}
where the coefficients $C_{i}$ $(i=1,2,\cdots, 9)$ are listed as follows,
\begin{align}
C_1 &= \frac{\sqrt{f}(h-\xi b_t^2)}{ h^{3/2}} \, , \nn \\
C_2 &= \frac{\sqrt{f}\xi b_t}{\sqrt{h}} \,, \nn \\ 
C_3 &= \sqrt{\frac{f}{h}}\left(\frac{(\xi-2\kappa)b_t'}{r}- \frac{2\xi b_t}{r}\right) \, ,  \nn \\
C_4 &= \frac{2\kappa}{\sqrt{fh}} \, ,\nn  \\
C_5 &= -2\kappa \sqrt{fh}  \, , \nn \\
C_6 &= -\frac{ (\ell -1) (\ell + 2)\sqrt{fh}}{r^2} \, ,  \nn \\
C_7 &= \frac{ (\ell -1) (\ell + 2)(h-\xi b_t^2)}{\sqrt{f h} h  r^2} \, ,  \nn \\
C_8 &= -\frac{ (\ell -1) (\ell + 2)2 \xi b_t }{r^2 \sqrt{fh}}  \, ,  \nn \\
C_9 &= -\frac{2\kappa\ell(\ell+1)}{r^2}\sqrt{\frac{h}{f}}-\frac{\xi \big(h(-2+2f+rf')+r f h' \big)}{r^2 \sqrt{fh}} \, .  \nn
\end{align}
%

\subsubsection{The $\ell \geq 2$ modes }\label{leq2}

In this subsection, we consider the odd parity modes with $\ell \geq 2$. Note that the perturbed action~\eqref{action2} does not contain $\dot{h}_0$ term, thus there are only two dynamical fields, $h_1$ and $\beta$. 
However, since Eq.~\eqref{action2} involves $h_0'$, $h_0$ cannot be solved directly. By introducing a Lagrangian multiplier $\chi (t,r)$, Eq.~\eqref{action2} yields
%
\begin{widetext}
\begin{eqnarray}\label{action3}
 I'_{\rm odd}  =  \int \td t \td r && \Bigg\{C_1\bigg[-\chi^2  + 2 \chi \left(\dot h_1 - h_0' + \frac{2}{r}h_0 + \frac{ C_2 \beta' + C_3 \beta}{C_1}\right)\bigg]    - \frac{(C_2 \beta' + C_3 \beta)^2}{C_1}  + C_4 \dot \beta^2 + C_5    \beta'^2 \cr
     ~~ \cr
      && +  \left(C_6 h_1^2 + C_7 h_0^2 + C_8 \beta h_0\right) + C_9 \beta^2 \Bigg\}  \, .
\end{eqnarray}
It is not difficult to find that the former action~\eqref{action2} can be recovered by varying $\chi(t,r)$ in the new action~\eqref{action3}. Performing variation on $I'_{\rm odd}$ with respect to $h_0$ and $h_1$ gives
\begin{eqnarray}
&& C_1 \dot \chi -  C_6 h_1 = 0\, , \cr
~\cr
&& 2C_1 (\chi' -\chi) - \frac{4}{r} \chi -  (2 C_7 h_0 + C_8 \beta)  = 0 \, .
\end{eqnarray}
Then $I'_{\rm odd}$ can be rewritten in the quadratic form 
%
\begin{eqnarray}
    I'_{\rm odd}  =  \int \td t \td r \left[-\frac{C_1^2}{C_6} \dot \chi^2 -\frac{C_1^2}{C_7}\chi'^2 + U_\chi \chi^2 + U_{\chi \beta}\chi \beta + C_4 \dot \beta^2 + \left(C_5 - \frac{C_2^2}{C_1}\right) \beta'^2 + U_\beta \beta^2 \right] \, ,
\end{eqnarray}
\end{widetext}
or in the matrix form
\begin{equation}
    I'_{\rm odd}= \int \td t \td r \left(\dot {\vv{\chi}}^\intercal \boldsymbol{ K} \dot{\vv{\chi}} + \vv{\chi}'^\intercal \boldsymbol{ G} \vv{\chi}' + \vv{\chi}^\intercal \boldsymbol{M} \vv{\chi}  \right)\, .
\end{equation}
Here $\vv{\chi}$ = $(\chi,\beta)^\intercal$ and $\boldsymbol{K}, \boldsymbol{G}, \boldsymbol{M}$ are $2 \times 2$  matrices
\begin{align}
    \boldsymbol{K} &=
    \begin{pmatrix}
         -\frac{C_1^2}{C_6} && 0  \\
         0 && C_4
    \end{pmatrix} \, , \\
     \boldsymbol{G} &=
    \begin{pmatrix}
         -\frac{C_1^2}{C_7} && 0  \\
         0 &&  C_5 - \frac{C_2^2}{C_1}
    \end{pmatrix}  \,, \\
    \boldsymbol{M} &=
    \begin{pmatrix}
         U_\chi && U_{\chi \beta} \\
         0 &&  U_\beta
    \end{pmatrix} \, ,
\end{align}
where
\begin{eqnarray}
&& U_\chi = -C_1 - \frac{6C_1^2 - r^2 C_1C_1''}{r^2 C_7} -\frac{r C_1 C_1' C_7' + 2 C_1^2 C_7' }{rC_7^2} \, , \cr
~\cr
&& U_\beta= C_9 + \left(\frac{C_2 C_3}{C_1}\right)' - \frac{C_3^2}{C_1} -\frac{C_8^2}{4 C_7}  \, , \cr
~\cr
&& U_{\chi \beta}=2C_3 -\frac{2C_1 C_8}{r C_7} + C_1 \left(\frac{C_8}{C_7}\right)' \, .
\end{eqnarray}
We next shall analyze the ghost and gradient (in)stability of $\vv{\chi}$ following the scheme introduced in Sec.~\ref{gins}. One can find that the coefficients of kinetic terms of $\dot \chi^2$ and $\dot \beta^2$,
\begin{align}
    C_4 &= \frac{2 \kappa}{\sqrt{f h}} > 0 \, , \\
    \frac{C_1^2}{C_6} &= -\sqrt{\frac{f}{h}} \frac{r^2 (h-\xi b_t)^2}{(\ell+2)(\ell-1)h^3} < 0 \,,
\end{align}
yielding 
\begin{equation}
 \boldsymbol{K}_{11}> 0 \, , \quad\quad \det \boldsymbol{K} > 0 \, . 
\end{equation}
It, therefore, shows that for the $\ell \geq 2$ case of the odd parity modes, the bumblebee BHs have \emph{no ghost instability}. 

In the following, we shall investigate the gradient instability of bumblebee BHs following the scheme introduced in Sec.~\ref{gins}. We assume that $\vv{\chi}$ has a wave solution along the radial direction,
\begin{equation}\label{planewave}
    \vv{\chi}= \vv{\chi}_k \te^{{ i}(\omega t - k r)} \, , 
\end{equation}
where $\vv{\chi}_k$ is a constant vector. We first consider the high-energy limit along the radial direction. 
In other words, we take limits $\omega \to \infty$ and $k \to \infty$,  but keeping $\omega/k$ a finite value. In order for $\vv{\chi}_k$ to have a nontrivial solution, we have
\begin{equation}\label{rad}
    \det \big(\omega^2 \boldsymbol{K}+ k^2 \boldsymbol{G} \big) = 0 \, .
\end{equation}
We then define the locally propagating speed of $\vv{\chi}$ along the radial direction,  
$$c_r \equiv \frac{\td r_*}{\td  \tau} = \frac{\omega}{k \sqrt{fh}} \,, $$
where the proper time $\tau \equiv \int  h \td t$ and the tortoise coordinate satisfies ${\td r_*}/{\td r} = f$. Then, Eq.~\eqref{rad} gives two solutions for the radial speed, 
\begin{equation}
    c_{r_{1}}^2 = \frac{C_2^2 - C_1 C_5}{C_1 C_4 f h} \, , \quad\quad  c_{r_{2}}^2 = -\frac{C_6}{C_7 f h} \,.
\end{equation}
Recall that the no-ghost condition implies $C_4 >0$ and $C_6 <0$. For the absence of gradient instability along the radial direction, namely 
\begin{equation}
    c_r^2 > 0 \,  ,
\end{equation}
we have the following constraints of gradient stability for bumblebee BHs,
\begin{eqnarray}\label{conr}
  \frac{C_2^2 - C_1 C_5}{C_1} =& \frac{\xi^2 b_t^2 + 2 \kappa(h-\xi b_t^2)}{h-\xi b_t^2} &> 0 \, , \cr
   ~\cr
     C_7 =& \frac{(\ell-1)(\ell+2)(h -\xi b_t^2)}{r^2 \sqrt{f}h^{3/2}}  &>0 \, .
\end{eqnarray}
It is easy to verify that when $h(r)>\xi b_t(r)^2$, namely that $c_{r_1}^2 >0$ is satisfied, $c_{r_2}^2 >0$ is also satisfied. 
Therefore, for bumblebee BHs, the condition for the absence of gradient instability associated with the radial direction is 
\begin{equation}\label{gracond}
    h(r) - \xi b_t(r)^2 > 0 \, .
\end{equation}

In addition to the radial direction, instability may also arise along the angular direction, depending on the propagating speed of $\vv{\chi}$. 
Now we consider the high-energy limit but along the angular direction. In other words, we take $\omega \to \infty$ and $\ell \to \infty$, but keep $\omega/\ell$ being finite. 
This limit indicates that the perturbation field, $\vv{\chi}$, propagates with large energy and angular momentum but the speed of propagation is finite. For nontrivial $\vv{\chi}_\ell$ solution, we have
\begin{equation}\label{ang}
    \det \big(\omega^2 \boldsymbol{K} + \boldsymbol{M} \big) = 0 \, .
\end{equation}
In the large-$\ell$ limit, the speed of propagation along the angular direction is defined as
$$c_\Omega \equiv r \frac{\td \theta}{\td \tau} = r\frac{\td \theta}{\sqrt{h}\td t} = \frac{r \omega}{\sqrt{h}\ell} \,.$$
With this definition and in the large-$\ell$ limit, $c_\Omega^2$ can be solved from Eq.~\eqref{ang} as
\begin{align}
    c_{\Omega_1}^2 &= \frac{h}{(h-\xi b_t^2)}\, , \\
    c_{\Omega_2}^2 &= \frac{\xi^2 b_t^2 + 2 \kappa(h-\xi b_t^2)}{2\kappa (h -\xi b_t^2)} \, .
\end{align}
We can see that the no tachyonic instability condition associated with the angular speed, namely,
\begin{equation}
c_{\Omega}^2 > 0 \,,
\end{equation}
shares the same constraints given by the no gradient instability condition associated with the radial speed in Eq.~\eqref{gracond}.

\subsubsection{The $\ell = 1$ modes }\label{leq1}

We have discussed the instabilities of bumblebee BHs for $\ell \geq 2$ odd-parity modes. 
Now we separately discuss the dipole mode, namely, $\ell = 1$. As we discussed previously, when $\ell =1$, the gauge degree of freedom is not fixed in Eq.~\eqref{acpert} since $h_{ab}$ identically vanishes. 

Recalling the gauge transformation rules given in Eq.~\eqref{trandiff}, one can simplify the gravitational perturbation by choosing a gauge with $h_1 = 0$, implying that
\begin{equation}
\Lambda(t,r) = -r^2 \int  \frac{h_1 (t,r')}{r'^2}  \td r' + r^2 {\cal F}(t) \, ,
\end{equation}
where ${\cal F}(t)$ is an arbitrary function with respect to $t$. Performing variation on the action~\eqref{action2} with respect to $\beta$ and $h_0$, and setting $h_1 = 0$, we have
\begin{equation}
\dot \Xi = 0 \, , \quad\quad  \Xi' = 0  \,,
\end{equation}
where
\begin{equation}
    \Xi \equiv 2 C_1 \left(h_0' - \frac{2}{r} h_0 \right) - 2( C_2 \beta' + C_3 \beta ) \, .
\end{equation}
It yields
\begin{equation}
    \Xi = {\cal C}_1 \, ,
\end{equation}
where ${\cal C}_1$ is an integration constant. Then $h_0$ can be solved as
\begin{equation}
    h_0 = r^2 {\cal F}_2 (t) + r^2 \int \td \tilde r \frac{{\cal C}_1 \tilde r + 2 C_4\beta + 2 C_3 \tilde r  \beta' }{2 C_1 \tilde r^3} \, ,
\end{equation}
where ${\cal F}_2 (t)$ is a gauge mode which can be eliminated by setting ${\cal F}(t) = \int \td t {\cal F}_2(t)$. 
In addition, if shutting down the vector perturbation with $\beta = 0$, namely that $h_0$ does not depend on time, then ${\cal C}_1$ is related to the angular momentum of a slowly rotating BH~\cite{Kobayashi:2012kh, Ogawa:2015pea}.

After integrating by parts, action~\eqref{action2} then reduces to
%
\begin{widetext}
\begin{eqnarray}\label{lag1}
   I'_{\rm odd} = && \int \td t \td r \left\{ C_4\dot \beta^2  - \left(C_5 - \frac{C_2^2}{C_1}\right) \beta'^2  
    + \left[C_9 + \left(\frac{C_2 C_3}{C_1}\right)' - \frac{C_3^2}{C_1}\right] \beta^2 \right\} \, , 
\end{eqnarray}
\end{widetext}
where we set ${\cal C}_1 = 0$ for convenience. The reduced action \eqref{lag1} shows that only one dynamical field, $\beta(t,r)$, propagates. It is obvious that for $\ell =1$, a bumblebee BH has {\emph no ghost instability} because
\begin{equation}
C_4 = \frac{2\kappa}{\sqrt{f h}} > 0 \, .
\end{equation}
Since  $\ell$ has been fixed, one can only read  $c_r^2$ in the high-energy limit along the radial direction and it is easy to find that the \emph{no gradient instability} condition is 
\begin{equation}
   C_5 - \frac{C_2^2}{C_1} = \frac{\xi^2 b_t^2 + 2 \kappa(h-\xi b_t^2)}{h-\xi b_t^2} >0 \, ,
\end{equation}
which is consistent with condition~\eqref{conr}. To summarize our results, for the odd-parity modes, the condition of \emph{no gradient instability} for  bumblebee BHs is
\begin{equation}\label{contot}
    h(r) - \xi b_t(r)^2 > 0 \, .
\end{equation}

In addition to the ghost instability, gradient instability, and tachyonic instability, instabilities associated with the quasinormal modes (QNMs) of BHs may arise when solving the equations of motion of $\vv{\chi}$ under appropriate boundary conditions~\cite{Konoplya:2011qq}. However, it is out of the scope of this work. We leave it to our future work to analyze the QNMs of  bumblebee BHs. 

\subsection{Numerical results}\label{sta}

In the previous section, we find that a bumblebee BH has no ghost instability but may have gradient instability and tachyonic instability, and obtain the condition~\eqref{contot} for avoiding them. In the following, we numerically investigate the condition and give constraints on the vector charge of bumblebee BHs. 

Based on Eq.~\eqref{contot}, a direct conclusion is that for  $\xi \leq 0$,
\begin{equation}
     h(r)-\xi b_t(r)^2>0 
\end{equation}
is always satisfied, indicating that these bumblebee BHs (including the RN BH) have no gradient instability nor tachyonic instability. 

For $\xi >0$, the condition becomes complicated so we shall discuss it case by case. We begin with  $\xi = 2\kappa$ since then the bumblebee BH has an analytical solution, the Schwarzschild metric with a nontrivial vector field given in Eq.~\eqref{specase2}. Equation~\eqref{contot} follows that
\begin{equation}
   h-\xi b_t^2=\frac{(r-2M) \big[2 MQ^2+r(M^2 - Q^2) \big]}{r^2 M^2} > 0 \, ,
\end{equation}
implying that when $M>Q$, $h-\xi b_t^2 $ is always positive, while $Q > M$, $h-\xi b_t^2 $ is positive only when $r$ is small. {\color{black} Therefore}, there must exist a critical radius,
\begin{equation}
    r_c = \frac{2 M Q^2}{Q^2 - M^2} \,,
\end{equation}
where $h-\xi b_t^2 $ changes the sign, implying that instabilities occur. If $Q = M$, $r_c$ is located at the infinity.  We thus easily draw a simple conclusion that when $\xi = 2\kappa$, the condition for no gradient instability nor tachyonic instability for a stealth Schwarzschild BH is 
\begin{equation}\label{MlQ}
    M \geq Q \, .
\end{equation}
It is a conclusion that, while concise, is rich in physical significance. It implies that even though stealth Schwarzschild BHs admit solutions with $Q>M$, the conditions of no instabilities indicate that such solutions cannot exist stably.  
Similar phenomena where Schwarzschild BHs with nontrivial vector field suffer the gradient instability also appear in the Einstein-aether theory and other vector-tensor theories~\cite{Tsujikawa:2021typ, Kase:2018voo}. 

\begin{figure}[h]
  \begin{center}
\includegraphics[width=0.48\textwidth, trim=0 30 0 5]{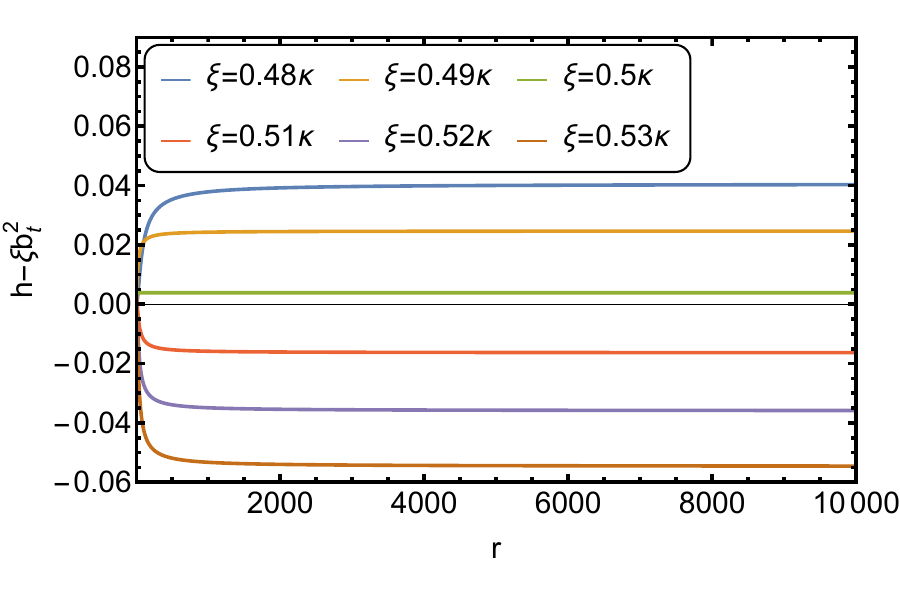}
  \caption{$h-\xi b_t^2$ against $r$ for various {\color{black} $\xi$ around $0.5\kappa$}. We verify that when {\color{black}$\xi \leq 0.5\kappa$}, $h-\xi b_t^2$ is a monotonically increasing function with respect to $r$.} 
  \label{xib}
  \end{center}
\end{figure}

For a general $\xi$, there is yet no analytical but numerical BH solutions. Following the hints given by the case of $\xi = 2 \kappa$, we thus choose a gradient stability indicator to show whether these numerical bumblebee BHs satisfy the condition~\eqref{contot}. Note that 
\begin{equation}
     h(r) \big|_{r \to \infty} = 1\, , \quad\quad    b_t(r) \big|_{r \to \infty} = \mu_{\infty} \, ,
\end{equation}
we thus choose the gradient stability indicator
\begin{equation}
  \left. h(r) - \xi b_t^2 \right|_{r \to \infty} =  1- \xi \mu_\infty^2 
\end{equation}
to show that when  $1- \xi \mu_\infty^2>0$,  bumblebee BHs have no gradient instability. Recall the example of $\xi = 2 \kappa$, we have
\begin{equation}
     1- \xi \mu_\infty^2 = \frac{M^2-Q^2}{M^2} 
\end{equation}
showing that the result of gradient stability for stealth Schwarzschild BHs is consistent with Eq.~\eqref{MlQ}. In addition to the case of $\xi = 2\kappa $, we find that $1-\xi \mu_\infty^2$ changes sign when {\color{black}
\begin{equation}
    \xi \approx 0.5 \kappa
\end{equation}
}
indicating that  bumblebee BHs begin to suffer gradient instability when {\color{black} $\xi > 0.5 \kappa$} (see  Table~\ref{Taxib}). 
\begin{table}[t]
    \centering
        \caption{Values of $|Q|/M$ and $1-\xi  \mu_\infty^2$ of bumblebee BHs for {\color{black} $\xi$ around $0.5 \kappa$}. It shows that $1-\xi  \mu_\infty^2$ changes sign when $\xi = 0.51\kappa$, indicating the critical point when the gradient instability happens. }
    \begin{tabular}{ccccccc}
    \hline\hline
        $\xi/\kappa$ & $0.48$   & $0.49$  & $0.50$  & $0.51$  & $0.52$  & $0.53$\\
    \hline
        $|Q|/[M]$ & $1.00016$   & $1.00227$  & $1.00193$  & $1.00187$  & $1.00216$  & $1.00287$\\
        $1 - \xi\mu_\infty^2$ & $0.040$ & $0.024$ & $0.0039$ & $-0.016$ & $-0.036$ & $-0.055$\\
    \hline
    \end{tabular}
    \label{Taxib}
\end{table}
Here we practically assume that when {\color{black} $\xi \leq 0.5\kappa$}, $h -\xi b_t^2$ is a monotonically increasing function with respect to $r$. Since when $r \to r_h$, one has $h-\xi b_t^2  \to 0$, such an assumption implies that $h-\xi b_t^2$ will always be positive for $r > r_h$ when  {\color{black}$\xi \leq 0.5\kappa$}. 
To verify it numerically, we show $h-\xi b_t^2$ against $r$ for various $\xi$ around $0.5\kappa$ in Fig.~\ref{xib}.
\begin{figure}[t]
  \begin{center}
  \includegraphics[width=0.45\textwidth, trim=10 10 0 0]{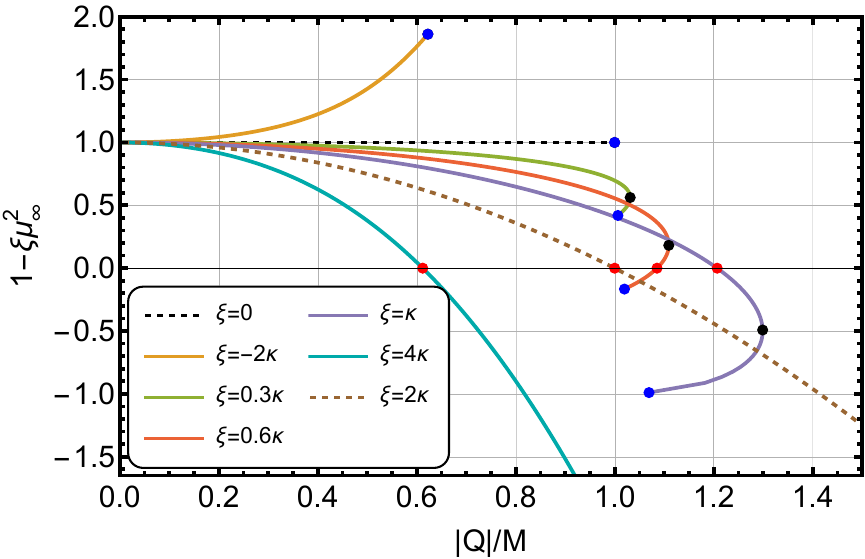}
  \caption{The relation between the gradient (in)stability indicator, $1-\xi \mu_\infty^2$, and the charge of bumblebee BHs.} \label{xiQ}
  \end{center}
\end{figure}

Recall that bumblebee BHs can be described by two parameters, $Q$ and $M$. We plot the gradient (in)stability indicator $1-\xi \mu_\infty^2$ with respect to $|Q|$ in Fig.~\ref{xiQ}. 
In the figure, we use the same $\xi$ as in  Fig.~\ref{rhQ}. When $\xi = -2\kappa$ and $0$, the gradient stability indicators are always positive for all $|Q|$, which is consistent with our previous statement that a bumblebee BH has no gradient instability when $\xi \leq 0$. 
Moreover, in Fig.~\ref{xiQ}, we denote the critical points where $1-\xi b_t^2$ changes sign with red dots and the maximum of $|Q|$ with black dots. 
For {\color{black} $\xi < 0.5\kappa$}, there does not exist any critical point (red dot), showing that a bumblebee BH does not suffer from the gradient instability {\color{black}(for example, see the green solid line for $\xi=0.3 \kappa $ in Fig.~\ref{xiQ})}.
We found that the critical points (red dots) appear when {\color{black} $\xi > 0.5\kappa$}. When $\xi$ increases but is less than $1.4\kappa$, there exist two bumblebee BH solutions for the same $|Q|$. When $\xi$ is close to but larger than {\color{black} $0.5\kappa$}, only one of the two BH solutions suffers gradient instability (for example, see the orange solid line for $\xi=0.6 \kappa $ in Fig.~\ref{xiQ}).
The critical points first move to the right, and then to the left after the critical points coincide with the maximum points (the black dots). When $0.75\kappa<\xi<1.4\kappa$, both of the two branches of bumblebee BHs suffer the gradient instability for larger $|Q|$ (see the {\color{black} purple} solid line for $\xi = \kappa$ in Fig.~\ref{xiQ} as an example). Furthermore, when $\xi > 1.4 \kappa$, there is only one bumblebee BH solution for one $|Q|$, and the bumblebee BH suffers the gradient instability when $1-\xi \mu^2_\infty$ is negative (see the case of $\xi = 2\kappa$ in Fig.~\ref{xiQ}).
\begin{figure*}[t]
  \begin{center}
\includegraphics[width=0.62\textwidth, trim=5 15 10 5]{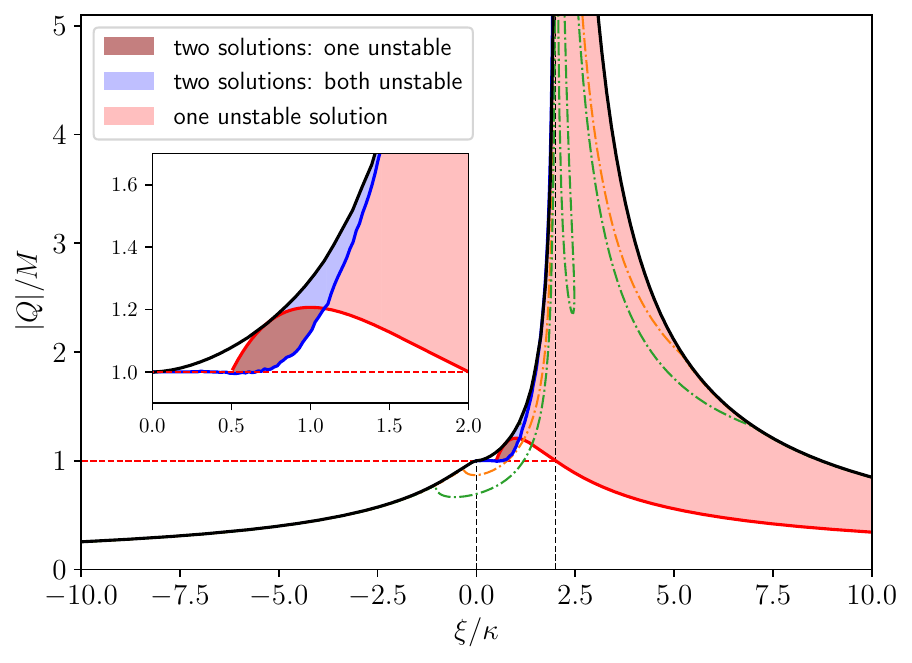}
  \caption{Constraints on the $\xi$-$|Q|$ plane from EHT observations and the gradient instability. {\color{black} The dark red region shows the parameter space where there are two branches of solutions but only one of them has gradient instability} while the blue region shows that both two branches of BHs have gradient instability. Furthermore, the light red region shows the parameter space where there is only one solution and it is unstable, with the red solid line giving the boundary for the gradient instability to happen. The orange and green dot-dashed lines show the bounds given by the observations of the shadows of supermassive BHs in M87 and Milky Way respectively~\cite{Xu:2023xqh}. The black lines give the boundary of the existence domain of BH solutions. The region enclosed by the black and blue lines indicates that there exist two branches of BHs for the same $|Q|$ and $\xi/\kappa$. } \label{total}
  \end{center}
\end{figure*}

As mentioned in the Introduction, \citet{Xu:2023xqh, Xu:2022frb} have put observational bounds on $|Q|$ associated with bumblebee BHs for various $\xi$ using the supermassive BH shadows of M87$^*$ and Sgr ${\rm A}^*$ from the EHT. Based on our results of the gradient stability for bumblebee BHs,  in Fig.~\ref{total} we give an extended plot of Fig.~\ref{xiQ} for more choices of $\xi$. The black line, formed by connecting the maximum $|Q|$ of bumblebee BHs, serves as the boundary of the BH solutions. The blue line serves as the boundary of the existence of two branches of solutions for the same $|Q|$ and $\xi$.
On the other hand, the green and orange dot-dashed lines are given by the observations of the shadows for the supermassive BHs in M87 and the Milky Way respectively. The red line, formed by connecting the critical points for different values of $\xi$ (some of which are denoted as red dots in Fig.~\ref{total}) serves as the boundary of the gradient stable and unstable regions. Therefore, the colored region represents bumblebee BHs suffering the gradient instability. 

In addition, due to the bound given from observations on shadows for the supermassive BHs, we find that when $\xi < 2\kappa$, even though the bumblebee BHs do not suffer the gradient instability, $|Q|$ of bumblebee BHs cannot be sufficiently large since such BH solutions do not exist. Specifically, when $\xi$ is negative, $|Q| < M$ is always satisfied. Nevertheless, when $\xi > 2 \kappa$, it admits bumblebee BHs carrying a considerable vector charge $|Q|$. However, the bumblebee BHs with $|Q|>M$ can rarely exist due to gradient instability.  
Therefore, we draw a conclusion that the bumblebee BHs could not carry considerable vector charge $|Q|$, and further when $\xi \geq 2\kappa$ or $\xi \leq 0$---equivalently $ \xi(\xi-2\kappa) > 0 $---the vector charge of a bumblebee BH cannot be larger than its mass,
\begin{equation}
  |Q| < M \, .
\end{equation}

{\color{black} It is worth mentioning that our stability analysis on the gradient of the bumblebee BHs, especially the onset of the gradient instability, is based on the plane-wave approximation, namely Eq.~\eqref{planewave}. For general solutions of $\chi$, the conditions of the (in)stability of the bumblebee BHs become more complicated. It is indeed a worthwhile topic and we will leave it for future studies. }

\section{Conclusions}\label{con}
In this paper, we have studied the dynamic (in)stability of bumblebee BHs in the bumblebee vector-tensor theory.  
Treating the BH spacetime as the background, we have considered perturbations of the gravitational field and the bumblebee field with odd parity and have investigated the associated (in)stability. {\color{black} Under the plane-wave approximation, our conclusions can be summarized as follows}.
\begin{enumerate}[(I)]
    \item Bumblebee BHs do not suffer the ghost instability.
    \item Bumblebee BHs suffer the gradient and tachyonic instabilities {\color{black} when $\xi > 0.5\kappa$ according to our numerical results}.
    \item The conditions for avoiding both the gradient and tachyonic instabilities for bumblebee BHs indicate a no-go theorem on the bumblebee charge. Specifically, when $\xi(\xi-2\kappa)\ge 0$, the vector charge of a bumblebee BH can not be larger than its mass.
\end{enumerate}

As shown in Fig.~\ref{total}, the theoretical condition for no gradient/tachyonic instability gives a stronger constraint on the charge of bumblebee BHs than the constraints given by the observed  BH images from  EHT when $\xi\gtrsim 1.3\ka$. In general, Fig.~\ref{total} also shows that if $\xi (\xi - 2\kappa) \geq 0$, bumblebee BHs with $|Q|>M$ can hardly exist in our Universe due to the gradient/tachyonic instability. This {\color{black} appears to be} a realization of the weak cosmic censorship conjecture in the bumblebee theory. In GR, for a RN BH, the weak cosmic censorship conjecture states that the charge cannot be larger than the mass, to avoid the naked singularity~\cite{Hawking:1970zqf, 1974IAUS6482P}.  
Our finding suggests that the gradient/tachyonic instability might be the physical mechanism behind the weak cosmic censorship, if we extend the Einstein-Maxwell theory to the bumblebee vector-tensor theory, potentially providing a physical mechanism for the weak cosmic censorship in a broader context of gravity theories. As the gradient/tachyonic instability also exists in other modified gravity theories, we expect there to be also such worthwhile constraints on the corresponding additional hairs of BHs in these theories. 
In addition to the ghost instability, gradient instability, and tachyonic instability of odd-parity perturbations considered in this work, instabilities related to the QNMs of bumblebee BHs, together with the even-parity perturbations and their stability properties, are also worthwhile topics for the stability analysis of bumblebee BHs. We leave them for future studies.

\begin{acknowledgments}
We thank Zexin Hu, Yingli Zhang, Jinbo Yang and Run-Qiu Yang for useful discussions.  This work
was supported by the National Natural Science Foundation of China (Grants~No.~12247128,
No.~11991053, No.~11975027, No.~11721303), the China Postdoctoral Science Foundation (No.~2021TQ0018, No.~2023M741999), the
National SKA Program of China (No.~2020SKA0120300), 
the Max Planck Partner Group
Program funded by the Max Planck Society, and the High-Performance Computing
Platform of Peking University. 
\end{acknowledgments}

\appendix

\section{The background field equations under the static spherical ansatz}
\label{app1}
Denoting $E_{\mu\nu} =G_{\mu\nu} - \kappa T^b_{\mu\nu}$, with static spherical ansatz the non-vanishing components of the background Einstein equations are
\begin{widetext}
\begin{eqnarray}
     0 = E_{tt} &=& \frac{h(1-f-rf')}{2r^2 \kappa} - \frac{1}{4}fb_t'^2 + \frac{\xi}{2\kappa} \Big(b_t'' + \frac{b_t'^2}{b_t} + \left(\frac{2}{r} + \frac{f'}{2f}- \frac{2h'}{h}\right)b_t' + \left(\frac{5h'^2}{4h^2} -\frac{h''}{h} - \frac{4fh'+rf'h'}{2 r fh}\right)b_t \Big)f b_t\, , \cr
     && ~\\
    0 = E_{rr} &=& \frac{h'}{2r \kappa h} + \frac{f-1}{2 r^2 \kappa f} + \frac{b_t'^2}{4h} + \frac{\xi}{8\kappa}\left(\frac{b_t^2 h'^2}{h^3}- \frac{2b_t b_t'h'}{h^2}\right) \, , \\
    ~\cr
    0 = E_{\theta \theta} &=& \frac{E_{\varphi \varphi}}{\sin^2 \theta} = \frac{r^2f}{4\kappa h} h'' + \frac{r(2f + r f')}{8\kappa h} h' - \frac{r^2 f}{8\kappa h^2}h'^2 + \frac{r f'}{4\kappa} - \frac{r^2 f}{4h}b_t'^2 + \frac{\xi}{8\kappa}\left(\frac{2 b_t'f h'}{h^2} - \frac{b_t f h'^2}{h^3}\right)r^2 b_t\, ,
\end{eqnarray}
where "$'$" denote the derivative with respect on $r$. The field equation of the background bumblebee vector has only the temporal component, namely
\begin{equation}
0 = b_t'' + \left(\frac{2}{r} + \frac{f'}{2f}-\frac{h'}{2h}\right)b_t' - \left(\frac{4 f h' + r f'h'}{4 r  f h} - \frac{h'^2}{4 h^2}+\frac{h''}{2h}\right)\frac{\xi}{\kappa} b_t \, .
\end{equation}
\end{widetext}
%

\bibliography{Bum_sta}

\end{document}